# Antiferromagnetic Potts Models on the Square Lattice

Sabino José Ferreira
*Departamento de Física – ICEx*
*Universidade Federal de Minas Gerais*
*Caixa Postal 702*
*Belo Horizonte, MG 30161 BRASIL*
Internet: SABINO@FISICA.UFMG.BR

Alan D. Sokal
*Department of Physics*
*New York University*
*4 Washington Place*
*New York, NY 10003 USA*
Internet: SOKAL@ACF4.NYU.EDU

May 16, 1994

### Abstract

We study the antiferromagnetic $q$-state Potts model on the square lattice for $q = 3$ and $q = 4$, using the Wang-Swendsen-Kotecký Monte Carlo algorithm and a new finite-size-scaling extrapolation method. For $q = 3$ we obtain good control up to correlation length $\xi \sim 5000$; the data are consistent with $\xi(\beta) = Ae^{2\beta}\beta(1 + a_1 e^{-\beta} + \ldots)$ as $\beta \to \infty$. For $q = 4$ the model is disordered even at zero temperature.

hep-lat/9405015  17 May 1994

The Potts model [1,2,3] plays an important role in the general theory of critical phenomena, especially in two dimensions [4,5], and has applications to various condensed-matter systems [2]. Ferromagnetic Potts models have been extensively studied over the last two decades, and much is known about their phase diagrams [2,3] and critical exponents [5,6]. But for antiferromagnetic Potts models, many basic questions remain open: Is there a phase transition at finite temperature, and if so, of what order? What is the nature of the low-temperature phase? If there is a critical point, what are the critical exponents and the universality classes?

In this Letter we report the results of a large-scale Monte Carlo study of the 3-state and 4-state antiferromagnetic Potts models on the square lattice, using the Wang-Swendsen-Kotecký (WSK) [7,8,9] cluster algorithm. We use a powerful new finite-size-scaling (FSS) extrapolation method [10,11] to estimate the infinite-volume correlation length $\xi$ and staggered susceptibility $\chi_{stagg}$. Using lattices up to $1536^2$, we can attain an accuracy of a few percent on $\xi$ and $\chi_{stagg}$ at correlation lengths $\xi$ as large as 5000. This allows us to conjecture the exact form of the critical behavior for the 3-state model.

The $q$-state Potts model is defined by the reduced Hamiltonian

$$\mathcal{H} = -J \sum_{\langle xy \rangle} \delta_{\sigma_x \sigma_y} , \qquad (1)$$

where the sum runs over all nearest-neighbor pairs of lattice sites, and each spin takes values $\sigma_x \in \{1, 2, \ldots, q\}$. The antiferromagnetic case corresponds to $J = -\beta < 0$.

Baxter [4,12] has determined the exact free energy (among other quantities) on two special curves in the $(J, q)$-plane:

$$e^J = 1 \pm \sqrt{q} \qquad (2)$$

$$e^J = -1 \pm \sqrt{4-q} \qquad (3)$$

Curve $(2_+)$ is known to correspond to the ferromagnetic critical point, and Baxter [12] conjectured that curve $(3_+)$ corresponds to the antiferromagnetic critical point. For $q = 2$ this gives the known exact value; for $q = 3$ it predicts a zero-temperature critical point ($J_c = -\infty$), in accordance with previous belief [13,14]; and for $q > 3$ it predicts that the putative critical point lies in the unphysical region ($e^{J_c} < 0$), so that the entire physical region $-\infty \leq J \leq 0$ lies in the disordered phase.

We remark that the $q = 3$ model is exactly soluble at zero temperature in an arbitrary magnetic field [14,15,16]. We would not be surprised if the model were found to be exactly soluble also at nonzero temperature (in zero field), at least in the sense of determining the exact asymptotic behavior as $\beta \to \infty$ (see [17,18,19] for some tantalizing ideas in this direction).

Nightingale and Schick [20], using a phenomenological-renormalization method based on infinite strips of width 2–8, confirmed the prediction of a zero-temperature critical point for $q = 3$, and claimed that the correlation length diverges as $\xi \sim \exp(c\beta^{\approx 1.3})$. Wang, Swendsen and Kotecký [7,8], using Monte Carlo, claimed to confirm this latter behavior. But this behavior seems *a priori* implausible to us:

the fundamental variable in the Potts model is $t = e^J$, so an ordinary power-law critical point $\xi \sim (t - t_c)^{-\nu}$ with $t_c = 0$ would correspond to $\xi \sim e^{\nu\beta}$. Moreover, we suspect that this model can be exactly solved, in which case $\nu$ would most likely be a rational number. We are unable to imagine any mechanism leading to $\xi \sim \exp(c\beta^\kappa)$ with $\kappa \neq 1$.

We simulated the models for $q = 3$ and $q = 4$. We measured the energy $E$, the staggered susceptibility $\chi_{stagg}$, and the (finite-volume) second-moment correlation length

$$\xi_L = \left( \frac{(\chi_{stagg}/F_{stagg}) - 1}{4\sin^2(\pi/L)} \right)^{1/2} \qquad (4)$$

where $\chi_{stagg} = \widetilde{G}(\pi, \pi)$ and $F_{stagg} = \widetilde{G}(\pi + 2\pi/L, \pi)$ [here $\widetilde{G}$ is the Fourier transform of the spin-spin correlation function].

For $q = 3$ we ran on $L \times L$ lattices with $L = 32, 64, 128, 256, 512, 1024, 1536$ at 153 different pairs $(\beta, L)$ in the range $\beta \leq 6.0$ (corresponding to $\xi \lesssim 20000$). Each run was between $2 \times 10^5$ and $2.2 \times 10^7$ iterations of the WSK algorithm, and the total CPU time was about 2 years on an IBM RS-6000/370. The WSK algorithm appears to have *no* critical slowing-down: we found $\tau_{int,\mathcal{M}^2_{stagg}} < 5$ uniformly in $\beta$ and $L$ [21]. We extrapolated to infinite volume by the method of [10,11], taking $\xi_{min} = 10$ and $L_{min} = 64$ and using a quartic fit for the FSS functions (see [11,22] for more details). We thereby obtained $\xi$ to an accuracy of about 1% (resp. 2%, 3%, 5%) at $\xi \approx 1000$ (resp. 2000, 5000, 10000). The errors on $\chi_{stagg}$ were roughly twice as big.

Our data are in clear agreement with the prediction of a critical point at $\beta = \infty$. The correlation length $\xi$ rises roughly like $e^{2\beta}$, and we initially thought that this was the exact asymptotic behavior. However, at $\beta \gtrsim 3.4$ ($\xi \gtrsim 75$), $\xi$ begins to rise *faster* than this (Fig. 1a). We therefore guessed a multiplicative logarithmic correction, i.e. $\xi \sim e^{2\beta}\beta^p$ for some power $p$; see Fig. 1b,c for $p = 1/2$ and $p = 1$, respectively. It is difficult to distinguish between $1/2 \lesssim p \lesssim 1$ without additional information on the corrections to the leading asymptotic behavior. We do not know how to carry out a low-temperature expansion around the (critical) zero-temperature state; but it is reasonable to expect that there exists an expansion in powers of $e^{-\beta}$, which corresponds to a minimum energy cost of one unit for an "overturned" spin. We therefore expect

$$\xi(\beta) = Ae^{2\beta}\beta^p \left[ 1 + a_1 e^{-\beta} + a_2 e^{-2\beta} + \ldots \right]. \qquad (5)$$

If we accept this Ansatz, a value $p \approx 1$ is clearly favored, with $A \approx 0.184$ and $a_1 \approx 11$ (Fig. 2).

The critical exponent $\nu = 2$ found here corresponds to an operator with scaling dimension $X = 2 - 1/\nu = 3/2$, which is one of the possibilities proposed by Saleur [18, p. 248]. A logarithmic correction $\beta^p \sim (\log t)^p$ with $p$ integer (particularly $p = 1$) can occur as a result of "resonance" between operators whose scaling dimensions are rationally related [23]. We hope that the numerical results presented here will serve as useful clues toward the exact solution of this model.

It is only fair to note that our data can also be fit reasonably well by $\xi \sim \exp(c\beta^\kappa)$ with $\kappa \approx 1.3$ (Fig. 3). It is on *theoretical* grounds that we prefer (5). We doubt that these two Ansätze can be reliably distinguished by numerical means in the foreseeable future.

The staggered susceptibility is consistent with the believed exact behavior [24] $\chi_{stagg} \sim \xi^{5/3}$, unmodified by any further powers of $\beta$. The energy per site is consistent with

$$E(\beta) = b_1 e^{-\beta} + b_2 e^{-2\beta} + \ldots \qquad (6)$$

with $b_1 \approx 0.22$ (Fig. 4).

For $q = 4$ the story is very brief: simulations on $L = 32, 64$ show that $\xi \lesssim 2$ uniformly as $\beta \to \infty$. Clearly there is no critical point in the physical region. Physically, there is so much entropy that the correlations decay exponentially even at zero temperature. This can be proven rigorously to occur on the square lattice for $q > 8$ [25,22], and our simulations confirm Baxter's [12] prediction that it occurs in fact for $q > 3$.

Details of this work will appear elsewhere [22].

We wish to thank Chris Henley, Paul Pearce, Andrea Pelissetto and Hubert Saleur for helpful comments. The authors' research was supported in part by the Conselho Nacional de Desenvolvimento Científico e Tecnológico–CNPq (S.J.F.), the Fundação de Amparo à Pesquisa do Estado de Minas Gerais (S.J.F. and A.D.S.), and U.S. National Science Foundation grant DMS-9200719 (A.D.S.).

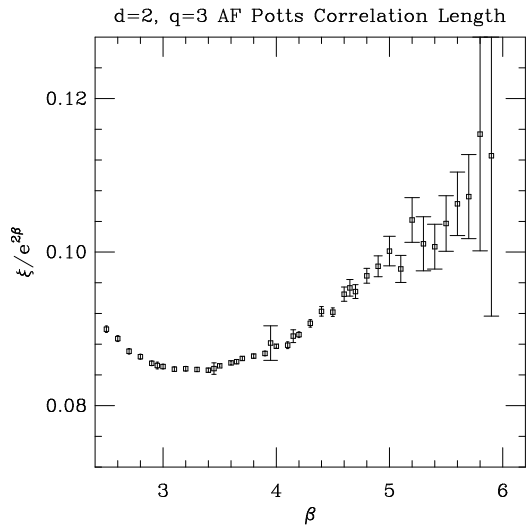

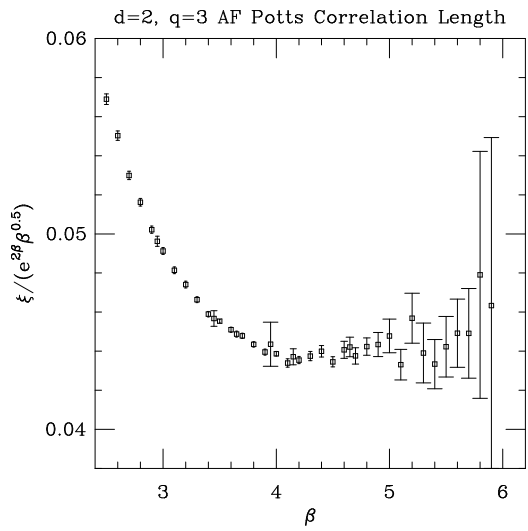

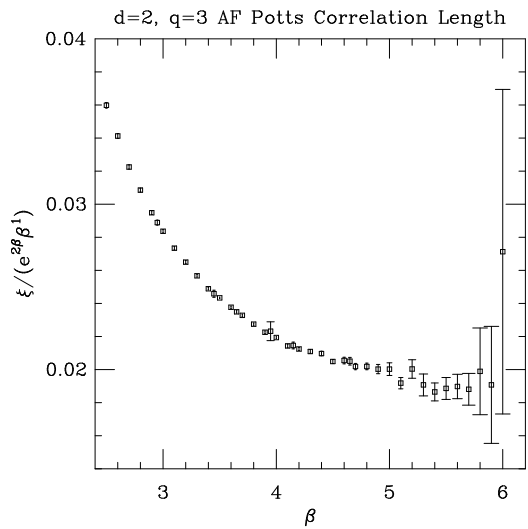

Figure 1: Infinite-volume correlation length $\xi$ divided by $e^{2\beta}\beta^p$ for (a) $p = 0$, (b) $p = 1/2$, (c) $p = 1$. Error bars are one standard deviation.

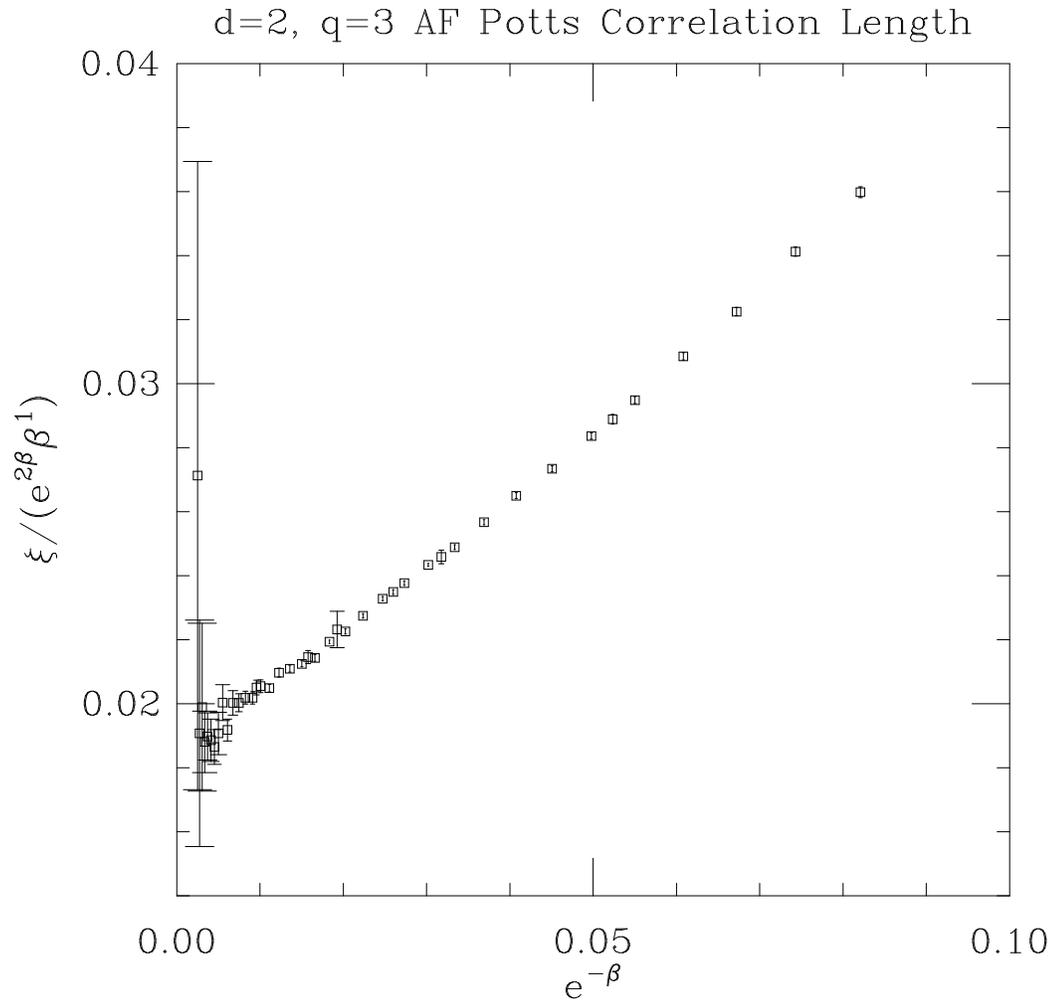

Figure 2: $\xi/(e^{2\beta}\beta^p)$ with $p = 1$, plotted versus $e^{-\beta}$. Note the nearly linear behavior, in good agreement with (5).

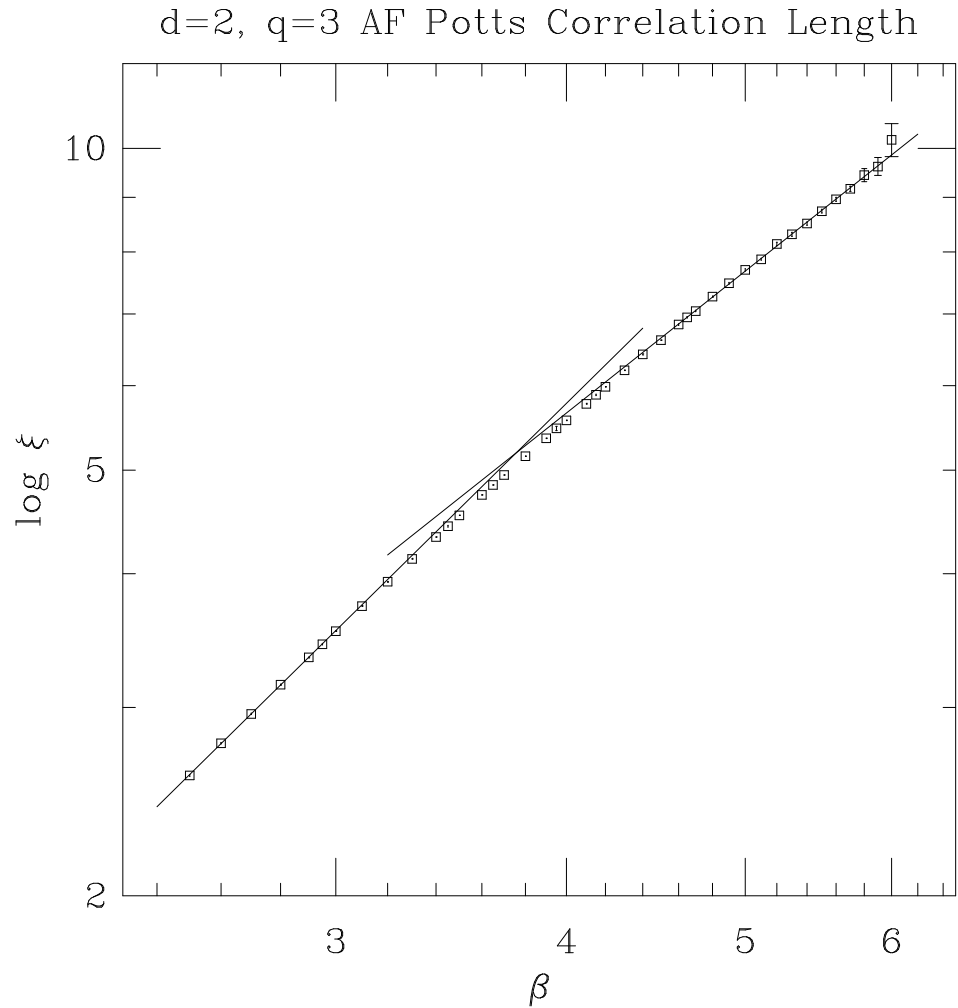

Figure 3: Log-log plot of $\log \xi$ versus $\beta$. The indicated asymptotes are, from left to right, $\xi = \exp(0.5469\beta^{1.7})$ and $\xi = \exp(0.8465\beta^{1.37})$.

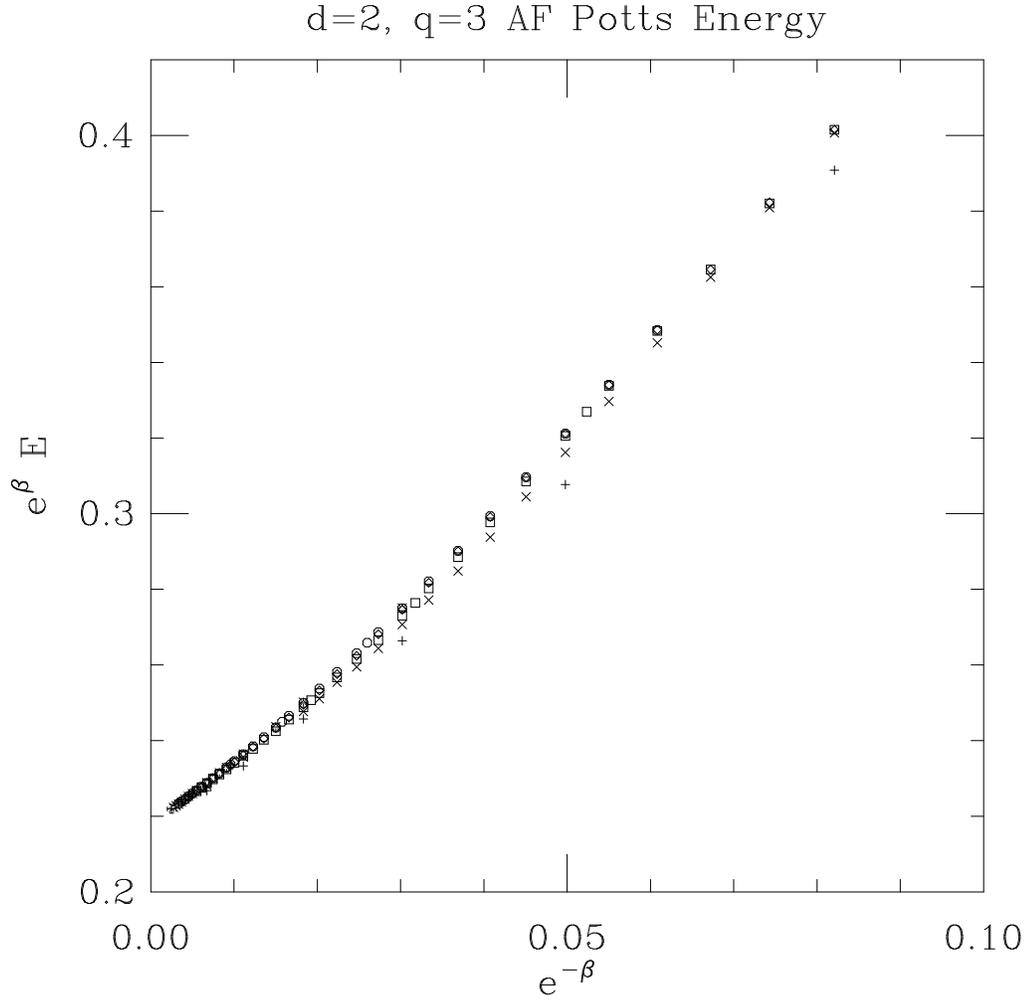

Figure 4: Energy per site $E$ divided by $e^{-\beta}$, plotted versus $e^{-\beta}$. Symbols indicate $L = 32$ (+), 64 (×), 128 (□), 256 (◇), 512 (○), 1024 (∗), 1536(⊞). Error bars are invisibly small. The uppermost points at each $\beta$ represent the infinite-volume limit. Note the nearly linear behavior, in good agreement with (6).